**Title:** Immunomodulatory role of an Ayurvedic formulation on imbalanced immune-metabolics during inflammatory responses of obesity and pre-diabetic disease.

**Authors:** Kamiya Tikoo*, Shashank Mishra*, V. Manivel, Kanury VS Rao*, Parul Tripathi*†‡ and Sachin Sharma*†‡

**Affiliations:** * International Centre for Genetic Engineering and Biotechnology, Aruna Asaf Ali Marg, New Delhi - 110067, India.

**ABSTRACT**

Obesity and related type 2 diabetes are associated with a low level chronic inflammatory state with disease status increasing at epidemic proportions worldwide. It is now universally accepted that the underlying immune-inflammatory responses mediated within the adipose tissue in obesity are central to the development of disease. Once initiated, chronic inflammation associated with obesity leads to the modulation of immune cell function. In the present study we aimed to investigate the effect of an ayurvedic formulation (named as Kal-1, an abbreviation derived from the procuring source) on diet-induced obesity and type II diabetes using C57BL/6J mice.
The study was planned into two groups using obese and pre-diabetic mouse model. The mice were fed on high-fat with increased diet and, different amounts (5, 20 and 75 µg) of Kal-1 were given with disease progression for 21 weeks in first group whereas mice were first put on the same diet for 21 weeks and then treated with same amounts of Kal-1. A significant reduction in body weight and fat pads, fasting blood glucose levels, insulin levels and biochemical parameters like triglycerides and cholesterol were observed in obese and diabetic mice+Kal-1 than control (lean) mice fed on low fat diet, though the optimum amounts of Kal-1 were 20 and 75 µg in first and second group respectively. A noteworthy immunological correction in important readouts *viz.* resistin, leptin, HMW adiponection and an array of pro- & anti-cytokines (IL-4, IL-10, IL-1α, IL-1β, IL-4, IL-6, IL-10, TNF-α and MCP-1) was also observed in both the groups with the mentioned amount of Kal-1. We conclude that Kal-1 has immunomodulatory potential for diet-induced obesity and associated metabolic disorders.

**Funding:**
This article is based on studies designed to assess the safety and potential physiologic effects of selected natural constituents of common Ayurveda mixtures. Studies were partially funded by PepsiCo, Inc and no additional external funding received for this study. The views expressed in this article are those of the author(s) and do not necessarily reflect the position or policy of PepsiCo, Inc. There are no other declarations relating to employment, consultancy, patents, products in development or marketed products.This does not alter our adherence to all the PLoS ONE policies on sharing data and materials.

**Competing interests:** The authors have declared that no competing interests exist. This does not alter our adherence to all the PLoS ONE policies on sharing data and materials.

†Corresponding Authors
‡ E-mail: parultripathi@rediffmail.com, sharmas009@gmail.com

**Introduction**

Globally, around 1.5 billion of the world's population are obese due to energy imbalance between calorie consumed and calorie expended [1]. Obesity is an inflammatory state particularly affecting the endocrine tissues mainly adipose tissues and, is referred as sub-clinical chronic mild inflammation, which is distinctive of clinical classic acute inflammation [2]. The inflammation embarks on adipocytes, which are major endocrine cells that have specialization in lipid storage, and with the expansion of adipocytes mass (hypertrophy and hyperplasia both), inflammation increases. The chronic low-grade inflammation activates the innate immune system that subsequently leads to insulin resistance and further leads to type II diabetes mellitus. The complete molecular and cellular inflammation between obesity and insulin resistance is just beginning to be revealed [3].

An important source of expansion or accumulation of fat in adipose tissue is consumption of high fat high sugar diets (HFHSD), fat accumulation is closely associated with qualitative changes in lipoproteins (low- and high-density), cholesterol and triglycerides. It is also suggested that free fatty acid production is increased during adipocyte mass expansion in obese state and might be playing an important role in blocking the insulin signal transduction [4].

Adipose tissue is of two different types – the white (WAT) and brown (BAT) adipose tissues; both can be clearly distinguished at morphological and functional level. BAT is known for heat production by thermogenesis, whereas WAT is considered as important endocrine tissue and contributes to the pathogenesis of insulin resistance and regulation of metabolic inflammation [5]. WAT (subcutaneous and epididymal) is a known site for storing calories as triglycerides and main site of inflammation related to obesity [6]. Along with inflammatory modulators such as leptin, resistin and adiponectin, a number of pro- (IL-1β, IL-6, IL-10 and TNF-α) and anti-inflammatory cytokines (IL-4 and IL-10) are also secreted by WAT. These adipokines dynamically affect metabolism as their production is considered to be regulated by nutritional state [7]. Anomalous production of mentioned adipocytokines and activation of inflammatory signaling pathways *viz.* Jun N-terminal kinase (JNK) and inhibitor of NF-κB kinase (IKK) are closely associated with chronic low-grade inflammation [8].

In modern era of pharmaceutical, most of the anti-obesity and anti-diabetic drugs are found inconsistently effective and also have their side-effects. An alternative form of medicine, herbal or ayurvedic (a typical antique and religious type of medicine predominantly practiced in Asia) formulations are now being considered worldwide because of their utmost least toxic nature and side effects compared to synthetic drugs. These herbal formulations are well known to treat metabolic disorders including obesity and diabetes. For instance, Shao *et al.* [9] described the role of curcumin as an anti-obesity and -diabetes herbal medicine in an organized manner. Still, there are very limited systematic studies on the effect of herbal formulations on metabolic immune balance in obese and diabetic individual.

This study was conducted to test one such formulation Kal-1, a polyherbal decoction of seven different natural ingredients (Supplementary Table 1), for its affect on the metabolic and immune balance in the mouse model of diet-induced obesity and diabetes. The procuring source has authorized both 1) the use of Kal-1 name in the manuscript and 2) the listing of specific ingredients of Kal-1 in the Supplementary Table 1. We evaluated the efficacy of Kal-1 as anti-obesity and anti-diabetic agent, in addition to its utility in controlling low-grade systemic inflammation and the overall energy equilibrium. We report here that, in addition to ameliorating the symptoms both of obesity and diabetes, Kal-1 administration also restored the normal balance of pro- versus anti-inflammatory cytokines thereby skewing the immune response to more of anti-inflammatory type. Importantly, this activity was evident in regimens that probed possible potential value of Kal-1 to be explored further for supplementing it from the nutrient or food perspective to control imbalanced immune responses and resulting metabolic disease entities.

**Materials and Methods**

**Preparation of Kal-1**

Kal-1 formulation is essentially a concoction of 8 different ingredients implicated to play protective role in inflammation. The latter is prepared using a methodology prescribed by the ancient Ayurvedic texts. Briefly, the different ingredients are washed, cleaned, dried and sieved to get a coarse powder. The latter is then steamed and boiled till the starting volume reduces to one eighth. This is followed by filtration and boiling till the decoction reaches $1/4^{th}$ of the volume from the $1^{st}$ filtrate (Figure 9S). The resulting filtrate is then used as Kal-1, which is a dark brown liquid with a peculiar rotten leaf like smell. The Kal-1 dose to be administered in mice was calculated using human to mice dose conversion formula as describes elsewhere [10].

**Animals, Diets and Experimental set-up**

All animal studies were carried out at BIONEEDS (laboratory animals & preclinical services), Bangalore, India, and approved by institutional animal ethics committee (IAEC). All experimental protocols were done as per applicable national and international guidelines. BIONEEDS is approved by committee for the purpose of control and supervision of experiments on animals (CPCSEA), Ministry of forests and environments, Government of India. Briefly, 3-4 weeks-old male C57BL/6J mice (7-9g) were housed (3-5 animals per cage) under standard conditions. All animals were initially put on two different diets {low fat diet (LFD, D12492) containing 10% kcal fat and high fat with increased sugar diet (HFHSD, D03062301) containing 60%kcal fat in pellet forms, procured from Research diet, NJ, USA}. The first two weeks without the formulation were considered as "acclimatization phase" wherein mice were given their respective diets before starting the formulation. The administration of Kal-1 was done in two different disease rescue (up to 21 weeks) and treatment (up to 30 weeks) experiments in which mice were divided into five groups – LFD control group, HFHSD control group and three differen amountf Kal-1 amounts (5, 20 and 75 µg) supplemented with high fat high sugar diet (HFHSD+Kal-1) test groups. In rescue experiment, all three amounts of Kal-1 were administered for further 18 weeks after acclimatization period, whereas Kal-1 amounts with the same amounts are administered only for 8 weeks after 18 weeks in treatment experiment. The feed intake was monitored daily, which included residual feed quantification as well and body weights were recorded twice a week.

**Tissues isolation and blood collection**

At an interval of three weeks i.e. wk 3, 6, 9, 12, 15, 18 and 21 over the entire experimental period, mice were kept on fasting for a period of 5-6 hours prior to blood collection and then anesthetized with ether in rescue experiment, whereas the same procedure was followed only at wk 26 and 30 in treatment experiment. White adipose tissues (epididymal and subcutaneous) fat depots were removed carefully in both the cases, thoroughly rinsed with phosphate buffer saline and weighed.

Blood was collected from retro orbital sinus for serum separation. Blood glucose levels were measured by using a glucometer (Roche diagnostics GmbH, Germany) at above mentioned time periods.

**Biochemical analysis**

The serum concentrations of low density lipids (LDL), high density lipids (HDL), total cholesterol and triglycerides were assayed enzymatically by using an automatic analyzer (ERBA, Automated random access clinical chemistry analyzer, EM260, Mannheim, Germany) with their respective kits. The serum insulin levels were measured using commercial available ELISA kit (ALPCO ultra sensitive mouse insulin kit, Salem, NH, USA) for mouse. Leptin and resistin levels were measured using commercial available radioimmunoassay kits (Quantikine, mouse Leptin and mouse resistin, Immunoassay, R & D, Minneapolis, MN, USA) and high molecular weight adiponectin (ALPCO, Adiponectin mouse total, HMW, Salem NH, USA).

**Cytokine measurement**

An array of seven cytokines (pro- and anti-inflammatory) was measured in serum of the experimental groups. Briefly, sera from the blood were used to detect the following cytokines namely IL-1α, IL-1β, IL-4, IL-6, IL-10, TNF-α and MCP-1 using the Luminex system (Liquichip 200, Luminex xMAP Technology, Valencia, CA, USA) as per manufacturer's protocols. Kits for multiplex analysis were obtained from Millipore, Billerica, MA, USA. All samples were run in triplicate.

**Statistics**

To check the significant difference between LF, HF group and HF supplemented with different amounts of Kal-1 group, we performed two sample t-tests with the hypothesis that two independent samples are coming from distributions with equal mean and unknown but equal variance. The null hypothesis is tested at 5% significance level for all. P value less than 0.05 was considered to be statistically significant. All the data are expressed as mean ± SEM (n=5, each group).

## Results

**Simultaneous Kal-1 administration mitigates the effects of HFHSD on systemic inflammation and metabolic dysfunction in mice**

Here, we evaluated the effect of LFD, HFHSD and Kal-1 supplemented HFHSD group in C57BL/6J mouse model up to 21 weeks. We monitored body weights, metabolic biochemical parameters and immunological readouts like hormones and cytokines.

To assess the potential effects of KAL-1 on body weights regulation, we observed body weights of animals fed on LFD and HFHSD for a period of twenty one weeks (a period including fifteen days of acclimatization). Wherein, we screened several amounts of Kal-1 ranging from very low (0.04 µg) to high amounts (300 µg). A total of 275 mice were grouped (5 mice/group) into eleven different groups *viz.* LF control, HFHS control and HFHS supplemented with nine different test amounts Kal-1. A clear-cut dose dependent effect of Kal-1 was observed over the entire experimental phase (week 3 to week 21) at all the amounts of Kal-1. However, this trend excluded the two higher amounts of 150 and 300 µg, which could be potentially toxic). Body weight at Kal-1 amounts of 5, 20, 38 and 75 µg were observed to be closely comparable to LF control group (Figure S1). Therefore, for all further experiments, we mainly focused on three Kal-1 amounts of 5, 20 and 75 µg.

Additionally, a significant difference in mean body weights of LF and HFHS control groups was observed (8.8 gm or 26.3%, $p<0.0005$; Figure 1A)-HFHSD fed animals were heavier than LFD fed animals at week 21. Though the body weights of HFHS+5µg Kal-1 were found to be higher than LF control group but this was not statistically significant (2.4 gm or 7.1%, $p>0.05$). This was also true for HFHS+20µg Kal-1 group (1.1 gm or 3%, $p>0.5$). The group at a Kal-1 dose of 75ug showed a completely different profile with differences in body weights being 7.2gm or 21.5% ($p<0.05$) (Figure 1A).

Further, in order to ensure that any changes or effects observed were true effects of Kal-1, gavage control alone were also put (wherein the mice were administered same volume of distilled water). There was no difference observed in the weight of the animals between the control groups and gavage control group (Figure S2).

Furthermore, similar effect of Kal-1 was also observed on the weights of fat pads i.e. WAT. At week 21, the mean of relative weights of WAT viz. epididymal and subcutaneous fat depots were significantly higher in HFHS control animals than LF control animals (4 gm or 66% and 5.6 gm or 47.8%, respectively). Same was true for epididymal fat pads as well where Kal-1 at 20ug exerted the same effect which was almost comparable with low fat control group (weight being 2gm for both groups). Weights of fat pads for Kal-1 at 5 and 75ug were 1 and 3 g, respectively. Whereas weight of subcutaneous fat pads of HFHS+Kal-1 5, 20 and 75µg test groups were found to be less than LF control group (Figure 1B).

**Amount of feed taken does not affect the weight of animals as confirmed by pair fed experiments**

Since gain in body weight was significantly higher in mice receiving HFHSD as compared to the group being fed on LFD. And the HFHSD+Kal-1 group was almost comparable with LFD control group. We measured the feed intake and body weights up to week 21 in all the groups including pair-fed group to demonstrate that the difference between LFD and HFHSD control groups was due to high carbohydrate with increased sucrose content in HFHSD, not due to more eating of HFHSD (Figure S3 and S4). Furthermore, to verify the effect of Kal-1 on body weights of mice, we monitored the *ad libitum* feed intake in HFHSD fed control mice and restricted the amount of feed to HFHSD + pair-fed control and HFHSD+pair-fed+Kal-1 test groups. Supplementary figure 4 reveals that body weights of HFHSD + pair-fed control and Kal-1 supplemented test groups gained weight (47.0 and 35.1 gm, respectively) almost similar to HFHSD control group and LFD control groups (47.9 and 35.7 g, respectively) at week 21. These findings indicate that HFHSD diet promoted increase in animal body mass and KAL-1 is effective in reducing body weight gain.

Furthermore there was no observable change in the core body temperature in either control or Kal-1 supplemented test group. Rectal temperature of all the animals from pair-feeding experiment were also recorded for all four groups from week 15 to week 21 and it was ranging from 94.2$^o$F to 98.2$^o$F (Figure S5), which was again normal.

**Kal-1 rectifies the metabolic imbalance in HFHSD fed mice**

**Fasting blood glucose and insulin profiles**

To determine that high calorie diet results in a shift in immune balance which leads to symptoms towards development of diabetes, blood glucose levels were also recorded. This was done after 5-6 hours of fasting for all 275 animals which were further grouped (11 groups) in the same manner as mentioned earlier (Figure S6).

In comparison to LFD fed animals, blood glucose levels were significantly elevated (19 mg/dL or 13.7%, $p<0.005$) in HFHSD fed animals, whereas the difference was very less (4.5 mg/dL or 2.8%) in LF control and HFHS+KAL-1 20µg test animals. A significant diminution (16 mg/dL or 11.5%, $p<0.05$ and 34 mg/dL or 24.6%, $p<0.005$) in blood glucose levels were noticed in HFHS+KAL-1 5µg and 75µg treated animals group in compare to LFD fed animals respectively at week 21 (Figure 2A).

Fasting serum insulin levels were also measured for LF, HFHS HFHS+Kal-1 5µg, HFHS+Kal-1 20µg and HFHS+Kal-1 75µg test groups. At week 21, serum insulin levels were 20 (0.2 ng/dL), 30 (0.3 ng/dL) and 40% (0.4 ng/dL) higher in HFHS+Kal-1 75µg, HFHS+Kal-1 75µg and HFHS group, respectively than LF animals. However only HFHS and LF controls significantly differed ($p<0.05$) from each other. On the contrary, the insulin levels of HFHS+Kal-1 75µg animals were 20% lesser (0.2 ng/dL) when compared with LF control group.

**Effect on HDL, LDL, Cholestrol and Triglycerides levels**

At week 15, non-significant elevated levels (10.7, 31.1, 24 and 16.5%) of fasting serum in HFHS control animals were noticed for HDL, LDL, cholesterol and triglycerides, respectively as compared to LF group. Similar to body weight and insulin profile, dose dependent effects of Kal-1 were perceived only in HDL, cholesterol and triglycerides levels. Here again Kal-1 dose of 20ug was found to be optimal dose, which brings back the levels of biochemical parameters in HFHSD fed animals closer to LFD control group. Minimal differences of 0.8 and 3.1% for HFHS+Kal-1 20 and 5µg, respectively as compared to LF control mice were observed.

Similarly, the difference between HFHS+20µg Kal-1 and LF control mice was less (2.5%) as compared to HFHS+Kal-1 5µg (18.7%). The divergence in the levels of triglycerides was almost equal (10 and 11.3%) between LF control group and HFHS+Kal-1 20 and 5µg test groups, respectively. Serum levels of HDL, cholesterol and triglycerides for HFHS+Kal-1 75µg test group were observed lesser than LF control group and these levels were not significant.

Kal-1 was also found to be effective on LDL profile with all three amounts; however the effect was not dose dependent. In contrast to HDL, cholesterol and triglycerides, the levels of LDL with HFHS+Kal-1 5 and 75µg were close to LF control rather than HFHS+Kal-1 20µg test group (Figure 2C). These observations were again not significant.

**Kal-1 corrects immunological readouts in HFHSD fed mice**

HFHSD altered all the immunological readouts viz hormones and cytokines pattern in HFHSD group in comparison to LFD group. These altered patterns came back to the normal when Kal-1 was administered in the similar groups of HFHSD group and these parameters were tracked over the same time phase as done earlier for body weights and biochemical readouts.

**Hormone and cytokine production is affected by HFHSD intake and Kal-1 administration**

We noticed dose dependent effect of Kal-1 for all three hormones *viz.* resistin, leptin and HMW adiponectin at week 15. The differences between LFD and HFHSD control groups were 50% (0.005), 56% ($p<0.0005$) and 38% (0.0001) for resistin, leptin and HMW adiponectin, respectively which were statistically significant (Figure 3A, 3B, 3C). For both leptin and resistin, regulation (though not significant) was exhibited at Kal-1 dose of 20ug which was in concordance with other previous observations as well. The levels of adiponectin, were also brought back close to LF control group at Kal-1 20µg test dose ($p<0.005$).

Two panels of cytokines namely pro- (IL-1α, IL-1β, IL-6, MCP-1 and TNF-α) and anti-inflammatory (IL-4 and IL-10) were analyzed in serum at week 15. Statistically significant differences were observed in LFD and HFHSD groups (33 pg/ml or 39%, p<0.0001 and 38 pg/ml or 41% p<0.0001) for IL-4 and IL-10 concentrations, respectively. These results suggest that increased body weight and related metabolic disorders due to HFHSD also affected the concentrations of anti-inflammatory cytokines.

Similar to IL-4 and IL-10, T-test showed that all studied pro-inflammatory cytokines were significantly different between LFD and HFHSD fed animals. The highly significant difference in values were ranging from 57 pg/ml or 42% (p<0.00001), 170 pg/ml or 80% (p<0.00001), 112 pg/ml or 70% (p<0.00001), 67 pg/ml or 53% (p<0.000008) and 226 pg/ml or 82% (p<0.000005) respectively for IL-1α and IL-1β, IL-6, MCP-1 and TNF-α. In contrast, the differences between LFD group and HFHS+Kal-1 20µg were parallel (p<0.0007) for all examined pro-inflammatory cytokines except MCP-1 and these two groups were closer to each other at 8%, 28%, 18% and 54% when compared to rest other amounts of Kal-1 i.e. 5 and 75µg. For MCP-1, LFD and two test amounts of Kal-1 (HFHS+Kal-1 20 & 75µg) were different with each other at the same extent i.e. at 34 pg/ml or 36%. However, HFHSD+Kal-1 5µg group was less different (19 pg/ml or 24%) with LFD group (Figure 4B).

Thus, both the readouts *viz.* hormones and cytokines expressions showed marked levels of correction in Kal-1 supplemented group.

**Kal-1 treatment restores the inflammatory balance in mice fed on HFHSD**
We also monitored the effect of Kal-1 post 18 weeks of obesity induction by feeding the mice on HFHSD, which was reverse of what we did earlier. Here also we observed the body weights, fasting blood glucose, blood biochemistry, serum hormones and cytokines in the same manner as described earlier with the same number of mice per group.
During this study, body weights of HFHSD fed animals were significantly increased (11.6 gm or 30%, (p<0.0001) as compared to LFD fed animals post 21 weeks obesity induction period (Figure 5A). Kal-1 at the three above mentioned amounts (5, 20 and 75µg) was then begun to be administered for next 8 weeks in the HFHSD group. We observed that 8 weeks short term dietary treatment of Kal-1 at 75µg significantly reduced (20%, p<0.005) the body weight of HFHSD fed animals instead of HFHSD+Kal-1 at 20µg test animals (11%) (Figure 5B), though the effect of Kal-1 was again dose dependent as seen earlier (Figure 5A, a, b).

**Treatment with Kal-1 modulates blood glucose and serum insulin levels in obesity induced mice by HFHSD:**
Fasting blood glucose and serum insulin levels were measured at week 30 in experimental mice to assess the effect of treatment with Kal-1 on these aspects. Fasting blood glucose value of HFHSD fed was significantly higher (25 mg/dL or 17%, p<0.005) than LFD fed mice at week 30, and the mean concentrations of fasting blood glucose in the Kal-1 75µg treated mice were significantly less (22 ng/dL, 15% p<0.005) than mice fed on HFHSD and rest two amounts of 5ug and 20ug of Kal-1 were more close to HFHSD fed mice as shown in Figure 5B. Moreover, the fasting serum insulin levels were increased > 2 fold in HFHS grouped mice than LF grouped mice, whereas these levels were unaltered in HFHS+Kal-1 75µg grouped mice. The levels of HFHS+Kal-1 5 and 20µg test group were also comparable with LF control group but not as closer as for HFHS+Kal-1 75µg test group (Figure 5C).

**Functional relevance of dietary treatment of Kal-1 on serum biochemistry**

Following the same approach, serum biochemistry of individual mice in all five the groups (control groups and Kal-1 treated groups) were also examined. All parameters like serum HDL, LDL, cholesterol and triglycerides were distinguishable between LF and HFHS control groups, although the differences in HDL and cholesterol levels were only significant. Kal-1 at 75µg was again found to be a perfect dose to treat long term induced obesity in HFHSD fed mice for all parameters except cholesterol. The values for HDL, LDL and triglycerides of HFHS+Kal-1 75µg test group were comparable (26, 8 and 17 mg/dL, respectively) with LF control group, in contrast, values of cholesterol for HFHS+Kal-1 5, 20 and 75µg test groups were comparable (8, 7 and 10 mg/dL, respectively) with HFHS control group.

**Anti-obesity affect of Kal-1 as assessed through measurement of hormones and cytokines secreted during disease state**

To further examine the treatment effect of Kal-1 on pro- and anti-inflammatory parameters, a set of two hormones and seven cytokines was analyzed in the serum of LF and HFHS controls and HFHSD fed mice supplemented with Kal-1. For serum leptin and HMW adiponectin levels, a significant alteration were detected in LF and HFHS control groups ($p<0.02$ and $p<0.05$ respectively) and, similarly in HFHS control group and HFHS+Kal-1 75µg test group ($p<0.02$ and $p<0.05$, respectively) at week 30 (Figure 7A, 7B).

Figure 8A revealed that serum levels of LF control mice were 40% ($p<0.01$) and 47% (0.005) higher than HFHS control mice in anti-inflammatory cytokines, IL-4 and IL-10, respectively. On the other hand, a 72% ($p<0.01$) and 65% ($p<0.001$) decrease in the levels of these cytokines was observed for HFHS+Kal-1 75µg group than LF control group, however the levels of both cytokines in HFHS+Kal-1 75µg group were found 53% and 33% lower respectively even than HFHS control group at week 30.

Whereas at week 30, HFHSD fed animals showed increase in the pro-inflammatory cytokines levels (Figure 8B), when compared with LFD fed animals ($p<0.005$ for all except MCP-1), while Kal-1 at 75µg resulted significant reductions of 70, 62, 60, 42 and 76% on the cytokines levels, respectively in the serum of HFHSD fed animals ($p<0.005$ for all except MCP-1, $p<0.05$ for MCP-1).

**Discussion**

The present study confirms immunoregulatory effect of Kal-1, an ayurvedic formulation suggestive of controlling obesity and diabetes. Kal-1 is basically a decoction of seven different ingredients (with synergistic properties) which we thought (based on information provided by the procuring source) could be useful in regulating heightened or disturbed immune response especially during chronic low-grade inflammatory conditions viz obesity and diabetes. We tested the above formulation in well established diet induced mice models (C57BL/6J strain of mice) using skewing of immune response from a pro- to anti-inflammatory as one of the key elementary readout. The effect of Kal-1 on body fat mass, adipose tissues (epididymal and subcutaneous) weights, blood biochemistry including blood glucose level and insulin profile and adipocytokines was monitored on these HFHSD induced experimental mice.

A number of *in vivo* studies have been shown the effects of low- and high-fat diets on body weights, blood glucose level and inflammatory markers [11, 12, 13, 14]. It is noteworthy that diets play an important role in inflammatory modulations; especially high carbohydrate diet directly contributes to fat mass expansion in adipose tissues and then leads to inflammation and insulin resistance [15]. In the present study, it is revealed that HFD with increased sucrose significantly elevates the body weights, blood glucose and serum insulin levels in mice than LFD fed mice over the observation period. Moreover, preliminary screening study of body weights in mice fed on HFHSD with the different dose amounts of Kal-1 was also performed; consequently Kal-1 dose dependent reduction in body weights was also observed (Figure S1). On the basis of body weights reduction profile, a single dose amount of Kal-1 *i.e.* 20 µg was optimized as body weights of animals on HFHSD along with this amount was equally to that of body weights of LF fed mice in rescue experiment (Figure 1A). Unexpectedly, same type of observations were also noticed with Kal-1 in treatment study (Figure 5A, a), however 75 µg, a higher amount of Kal-1 was the optimal dose (Figure 5A, b). In the same manner, a decrease in the weights of epididymal and subcutaneous fat pads were also exemplified by Kal-1 (Figure 1B, a, b).

With these observational facts, it may be speculated that Kal-1 is effective either at the level of regulating adipocyte hypertrophy or on adipogenesis or both. Second, alteration in the fatty acids present in HFHSD from monounsaturated fatty acid to saturated fatty acid due to Kal-1 is another possibility as circulating saturated fatty acids plays a key role in obesity [16]. It must, however, be taken into account that HFHSD comprised with almost equal amount of both saturated as well as monounsaturated fatty acids (details not given).

Despite the fact that chronic low grade inflammation is directly linked with the consumption of high carbohydrate diets [17], sucrose is one of the important elements in HFD which is leading cause of obesity, high blood sugar and insulin resistance [18]. In this respect, expected higher levels of blood glucose and serum insulin levels were observed in obese mice control group (fed on HFHSD) than in lean mice control group (fed on LFD). Irrespective of the HFHS constituents in diet, significantly lower levels of glucose and insulin were observed in both test groups administered with Kal-1 20 and 75 µg (Figure 2A, 2B, 5B, b, 5C), which could not be explained

In continuation, one metabolically active hormone is resistin which is secreted by adipocytes, may contribute to obesity, insulin resistance, and diabetes in mice. In parallel with Steppan and Lazer findings [19], here we show that serum resistin levels of lean mice were lowered up to more than 2 fold compared to obese mice. In accordance to glucose and insulin profile, decreased levels of resistin were observed in obese mice exposed to Kal-1 and were comparable to lean mice (figure 3A). Therefore, it can be speculated that Kal-1 exhibits similar effects on resistin levels as is observed for blood glucose and serum insulin levels [19, 20].

Further, leptin is an adipokine, also secreted by adipocytes, considered a key pro-inflammatory cytokine. In addition to regulating food intake and energy homeostasis, this bio-active molecule also plays a potent role in modulating the immune response and inflammatory processes. Leptin is present in serum in direct proportion to the amount of adipose tissue therefore sum of energy in adipose tissue reveal the level of leptin in serum i.e. more energy: more production of leptin. Similar to previous explanation [21], amount of energy stored in adipocytes of obese mice fed the HFHSD were higher than in mice fed the LFD as leptin levels were found to be significantly more in obese mice than lean mice in current study. In corroboration with previous studies, increased leptin production can be positively correlated with adipocyte hypertrophy and hyperplasia [22, 23]. In both the experiments, the serum from mice on HFHSD supplemented with Kal-1 showed that leptin levels came back to the normal levels almost comparable with the levels seen in the LF diet group (Figure 3A, 7A). One possible explanation could be that Kal-1 stimulates and catalyses lipolysis and at the same time also regulates excess accumulation of fat cells in the body.

Unlike leptin, adiponectin, an adipocyte-specific secretary protein, is well-known for its anti-inflammatory action [24]. Shklyaev S *et al.* reported that adiponectin with sustained peripheral expression can improve insulin sensitivity too [25]. The hormone also contributes to the induction to produce anti-inflammatory cytokines and suppresses the pro-inflammatory cytokines [26]. Serum adiponectin levels decreases with obesity or with adiposity increases; though, the mechanism behind this reduction is still unclear. Similar to above mentioned fact, the adiponectin concentration shrink with body weight reduction after the administration of Kal-1 in rescue and treatment studies in the present investigation (Figure 3C, 7B). Consequently, it can be hypothesized that Kal-1 might be a contributing factor for a reduction in body weight.

Production and regulation of adipocytokines from adipocytes has been shown to be completely based on dietary conditions as dietary fats are directly associated with obesity and related metabolic disorders like diabetes. LF diet accompanies with decreased inflammatory markers whereas HFHS diet improved levels of pro-inflammatory cytokines [27]. In case of obesity and impaired glucose metabolism, chronic low-grade inflammation is considered to be a principal mechanism. The chronic inflammation can be only controlled by equilibrium between pro-inflammatory and anti-inflammatory cytokines.

Though, a number of adipocytokines are expressed in and secreted by adipocytes in which IL-1α, IL-1β, IL-6, TNF-α and MCP-1 are considered as classical pro-inflammatory cytokines in chronic inflammatory responses. It has been implicated by earlier studies that these cytokines are involved in the low-grade inflammation, impaired glucose metabolism and insulin resistance [28, 29, 30, 31, 32]. In accordance to previous studies, it is revealed in the present study that mentioned pro-inflammatory cytokines concentrations are higher in HFHSD fed animals in comparison to LFD fed animals. The serum levels of TNF-α and IL-6, key pro-inflammatory cytokines, are frequently increased in the obese state which is well in concurrence with the present study. tTNF-α actively participate in the development of insulin resistance and IL-6 is linked with type II diabetes. IL-1α (cell-associated molecule) and IL-1β (secretary protein) are members of IL family and recognized as immunomodulatory proteins. Both are related with obesity while IL-1β is also linked with obesity-induced diabetes [6]. Yu R et al. reported that one of the important pro-inflammatory cytokine is MCP-1 whose circulating levels were high in obese mice model [30]. Contrary to this and accordance to our study, decreased serum concentrations of two important adipokines IL-4 and IL-10 were found in obese animals compared to lean animals. This finding is in support with the fact that these adipokines have long been considered as anti-inflammatory cytokines [33, 34].

In the rescue experiment, HFHSD supplemented with 20 µg of Kal-1 suppresses obesity and related pro-inflammatory responses like insulin response and blood glucose levels by reducing levels of IL-1α, IL-1β, IL-6, TNF-α and MCP-1 and simultaneously elevating levels of anti-inflammatory IL-4 and IL-10 at week 15, however, here Kal-1 working amount was lower i.e. 5 µg (Figure 4A, 4B). On the other hand in treatment experiment, same types of profile were observed for both pro- and anti inflammatory adipokines. The only difference was in the amount of Kal-1, here it was 75 µg as observed in other biochemical parameters in treatment study. The possible explanation for this higher amount of Kal-1 is that herbal formulation with several ingredients like Kal-1 is required in little amount if this is administered with disease progression. Once disease is old and induced for a certain period of time, more than 2 fold of herbal formulation is required. To the best of our knowledge, this is the first study in which these both aspects of disease is covered simultaneously and single herbal formulation, Kal-1, controls chronic low-grade inflammation and maintain a balance between pro- and anti-inflammatory cytokines too by its immunoregulatory effect and then contributes to control weight gain and related metabolic problems.

In conclusion, our investigations imply that Kal-1 exhibit substantial anti-obesity concomitant with metabolic regulatory effect especially in terms of chronic low-grade inflammation, energy equilibrium and linked significant disorders. However, as mentioned above that Kal-1 constitutes with seven different herbal ingredients, it is very difficult to conclude that a combination or a single ingredient responsible for these encouraging responses. Indeed, clinical trials are needed in order to understand the relevance of formulation. It is also clear that diet rich in carbohydrate with increased sugar may affect on body weight, blood biochemistry including glucose and serum insulin, as well as the levels of inflammatory markers both pro and anti, and most importantly energy balance. It would also be fascinating to investigate Kal-1 mechanism and influenceonmetabolic process and pathways at transcriptional level.

**Authors Contributions**

Conceived and designed the experiments: KVSR PT SS. Performed the experiments: KT SM PT SS. Analyzed the data: KVSR PT SS. Wrote the paper: KVSR PT SS.

**Figure Legends**

**Figure 1.** Body and tissue weights in low fat diet fed control were higher than high fat high sugar diet fed control and Kal-1 doses rescue mice fed on high fat high sugar diets from being obese. **(1A)** Week wise effect of Kal-1 on body weights in high fat high sugar fed mice. **(1B)** Effect of Kal-1 on tissue weights in high fat high sugar fed mice at week 21. **a)** epididymal fat **b)** subcutaneous fat.
All doses (5, 20 and 75 µg) of Kal-1 were supplemented along with HFHSD. Here, the abbreviations mean: LF: Low fat control, HF: High fat high sugar control. All the values represent mean ± SEM from five animals.

**Figure 2.** Kal-1 rectifies the metabolic imbalance in mice fed on high fat high sugar diets. **(2A)** Effect of Kal-1 on fasting blood glucose levels in high fat high sugar fed diet at week 21. **(2B)** Effect of Kal-1 on fasting insulin levels in high fat high sugar fed diet at week 21. **(2C)** Effect of Kal-1 on various biochemical parameters (fasting) in high fat high sugar fed diet at week 15. **a)** HDL **b)** LDL **c)** Cholesterol **d)** Triglycerides.
All doses (5, 20 and 75 µg) of Kal-1 were supplemented along with HFHSD. Here, the abbreviations mean: LF: Low fat control, HF: High fat high sugar control. All the values represent mean ± SEM from five animals.

**Figure 3.** Kal-1 rectifies the hormonal imbalance in mice fed on high fat high sugar diets. **(3A)** Effect of Kal-1 on resistin levels in high fat high sugar fed mice at week 15. **(3B)** Effect of Kal-1 on leptin levels in high fat high sugar fed mice at week 15. **(3C)** Effect of Kal-1 on high molecular weight adiponectin levels in high fat high sugar fed mice at week 15.
All doses (5, 20 and 75 µg) of Kal-1 were supplemented along with HFHSD. Here, the abbreviations mean: LF: Low fat control, HF: High fat high sugar control. All the values represent mean ± SEM from five animals.

**Figure 4.** Kal-1 rectifies the inflammatory cytokines imbalance in mice fed on high fat high sugar diets. **(4A)** Effect of Kal-1 on anti-inflammatory cytokines in high fat high sugar fed mice at week 15. **(4B)** Effect of Kal-1 on pro-inflammatory cytokines in high fat high sugar fed mice at week 15.
All doses (5, 20 and 75 µg) of Kal-1 were supplemented along with HFHSD. Here, the abbreviations mean: LF: Low fat control, HF: High fat high sugar control. All the values represent mean ± SEM from five animals.

**Figure 5.** Body weights, fasting blood glucose and insulin levels in high fat high sugar fed control were lower than low fat fed control mice and Kal-1 treatment restore the body weights and blood glucose levels

successfully. After obesity and diabetes induction period up to 21 weeks, Kal-1 treatment was started from week 22 - 33. **(5A) a)** Effect of Kal-1 treatment on body weights in high fat high sugar diet fed mice at week 22, 26 and 30. Treatment with all doses (5, 20 and 75 µg) of Kal-1 was started only after 21 week (induction period) along with high fat high sugar diet. **b)** Effect of Kal-1 treatment with optimum dose (75µg) on body weights in high fat high sugar diet fed mice only at week 30. **(5B) a)** Effect of Kal-1 treatment on fasting blood glucose levels in high fat high sugar diet fed mice at week 22, 26 and 30. Treatment with all doses (5, 20 and 75 µg) of Kal-1 was started only after 21 week (induction period) along with high fat high sugar diet. **b)** Effect of Kal-1 treatment with optimum dose (75µg) on fasting blood glucose in high fat high sugar diet fed mice only at week 30. **(5C)** Effect of Kal-1 treatment on fasting insulin levels in high fat high sugar diet fed mice at week 30. Treatment with all doses (5, 20 and 75 µg) of Kal-1 was started only after 21 week (induction period) along with high fat high sugar diet.

Here, the abbreviations mean: LF: Low fat control, HF: High fat high sugar control. All the values represent mean ± SEM from five animals.

**Figure 6.** Kal-1 treats the irregularities in blood biochemical parameters in mice due to fed on high fat high sugar diets at week 30. Effect of Kal-1 treatment on different biochemical parameters viz. **a)** HDL **b)** LDL **c)** Cholesterol **d)** Triglycerides in high fat high sugar diet fed mice at week 30. Treatment with all doses (5, 20 and 75 µg) of Kal-1 was started only after 21 week (induction period) along with high fat high sugar diet.

Here, the abbreviations mean: LF: Low fat control, HF: High fat high sugar control. All the values represent mean ± SEM from five animals.

**Figure 7.** Treatment of Kal-1 brings back the levels of pro- and anti-hormonal levels to normal which were showing metabolic disturbed contour in high fat high sugar fed mice at week 22, 26 and 30. **(7A)** Effect of Kal-1 treatment on leptin profile in high fat high sugar diet fed mice at week 22, 26 and 30. **(7B)** Effect of Kal-1 treatment on high molecular weight adiponectin profile in high fat high sugar diet fed mice at week 22, 26 and 30. Treatment with all doses (5, 20 and 75 µg) of Kal-1 was started only after 21 week (induction period) along with high fat high sugar diet.

Here, the abbreviations mean: LF: Low fat control, HF: High fat high sugar control. All the values represent mean ± SEM from five animals.

**Figure 8.** Treatment of Kal-1 modulates the anti- and pro-inflammatory cytokines levels to normal which were showing abnormal pattern in high fat high sugar fed mice at week 22, 26 and 30. Treatment with all doses (5, 20 and 75 µg) of Kal-1 was started only after 21 week (induction period) along with high fat high sugar diet.

Here, the abbreviations mean: LF: Low fat control, HF: High fat high sugar control. All the values represent mean ± SEM from five animals.

**Supporting information**

**Figure S1.** Dose dependent effect of Kal-1 on mean body weights of mice fed on HFHSD at week 21. All doses (0.04-300µg) of Kal-1 were supplemented along with HFHSD. Here, the abbreviations mean: LF: Low fat control, HF: High fat high sugar control. All the values represent mean ± SEM from five animals.

**Figure S2.** Comparison of LF control and HFHSD control with their respective gavage control groups at week 21. Here, the abbreviations mean: LF: Low fat control, HF: High fat high sugar control. All the values represent mean ± SEM from five animals.

**Figure S3.** Feed consumption in pair feeding experiment **a)** at week 12 **b)** at week 21. Amount of Kal-1 was 2µg/gm body weight of mice. Here, the abbreviations mean: LF: Low fat control, HF: High fat high sugar control, PF: Pair-fed. All the values represent mean ± SEM from five animals.

**Figure S4.** Body weights in pair feeding experiment from week 15, 17, 19 and 21. Amount of Kal-1 was 2µg/gm body weight of mice. Here, the abbreviations mean: LF: Low fat control, HF: High fat high sugar control, PF: Pair-fed. All the values represent mean ± SEM from five animals.

**Figure S5.** Rectal temperature profile from week 15, 17, 19 and 21. Amount of Kal-1 was 2µg/gm body weight of mice. Here, the abbreviations mean: LF: Low fat control, HF: High fat high sugar control, PF: Pair-fed. All the values represent mean ± SEM from five animals.

**Figure S6.** Week wise effect of Kal-1 on fasting blood glucose in high fat high sugar fed mice. All doses ranging from 0.04µg to 75µg of Kal-1 were supplemented along with HFHSD. Here, the abbreviations mean: LF: Low fat control, HF: High fat high sugar control. All the values represent mean ± SEM from five animals.

**Figure S7.** Effect of Kal-1 treatment on different biochemical parameters *viz.* **a)** HDL **b)** LDL **c)** Cholesterol **d)** Triglycerides in high fat high sugar diet fed mice at week 21 and 26. Treatment with all doses (5, 20 and 75 µg) of Kal-1 was started only after 21 weeks (induction period) along with HFHSD. Here, the abbreviations mean: LF: Low fat control, HF: High fat high sugar control. All the values represent mean ± SEM from five animals.

**Figure S8.** Effect of Kal-1 treatment on fasting insulin levels in HFHSD fed mice **a)** at week 21 **b)** at week 26. Treatment with all doses (5, 20 and 75 µg) of Kal-1 was started only after 21 week (induction period) along with high fat high sugar diet. Here, the abbreviations mean: LF: Low fat control, HF: High fat high sugar control. All the values represent mean ± SEM from five animals.

**Figure S9.** Preparation procedure of Kal-1

**Table 1. Composition of Kal-1**

| S. No. | Sanskrit Name | Botanical Name | Part |
|---|---|---|---|
| 1. | Abhaya | Terminalia chebula | Fruit |
| 2. | Vibheetaki | Terminalia belerica | Fruit |
| 3. | Amalaki | Embelica officinalis | Fruit |
| 4. | Haridra | Curcuma longa | Rhizome |
| 5. | Vijaysar | Pterocarpus marsupium | Heartwood |
| 6. | Raktachitraka | Plumbago indica | Root |
| 7. | Khadira | Acacia catechu | Bark |

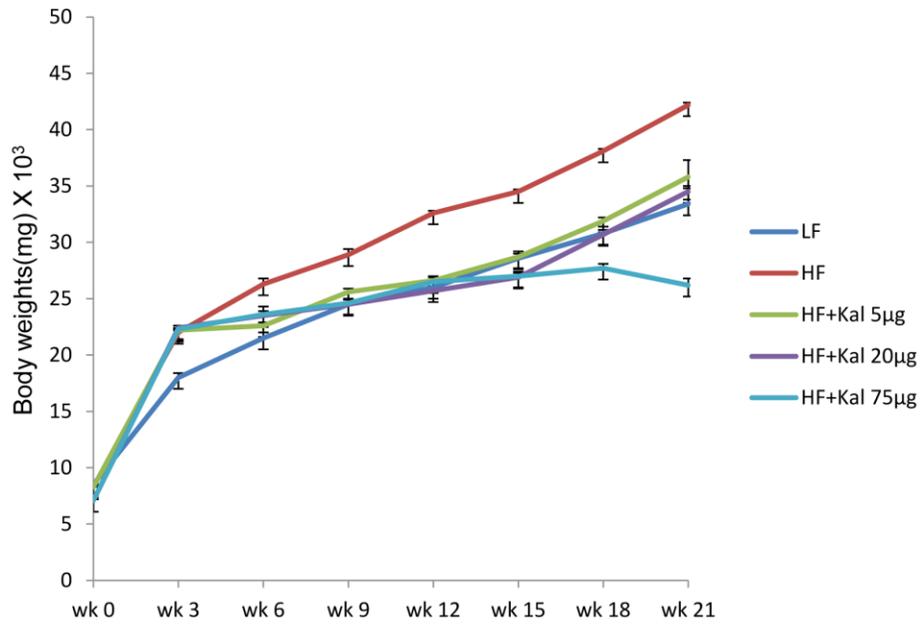

Figure 1A

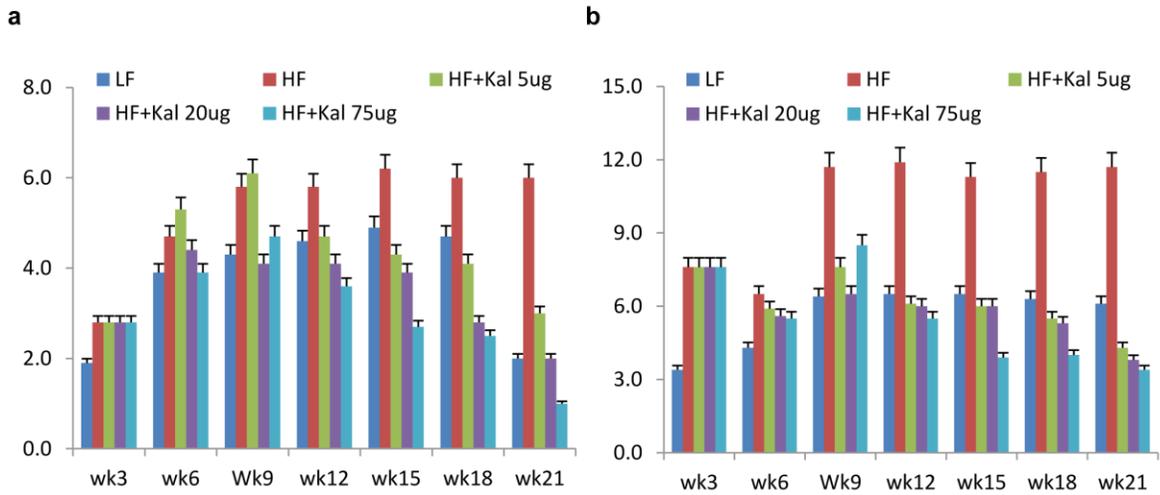

Figure 1B

**Figure 1**

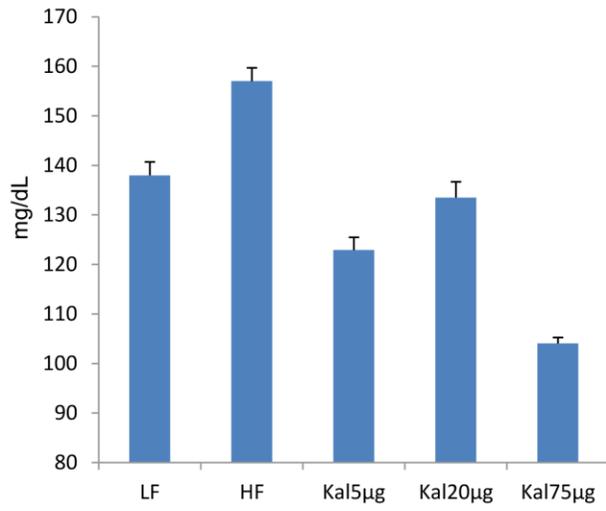
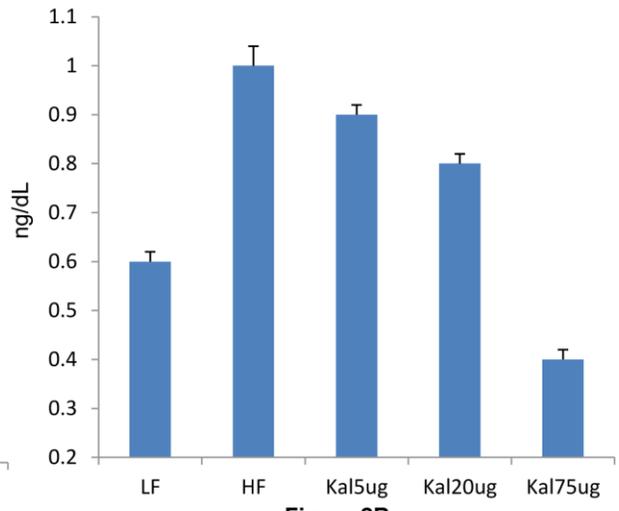
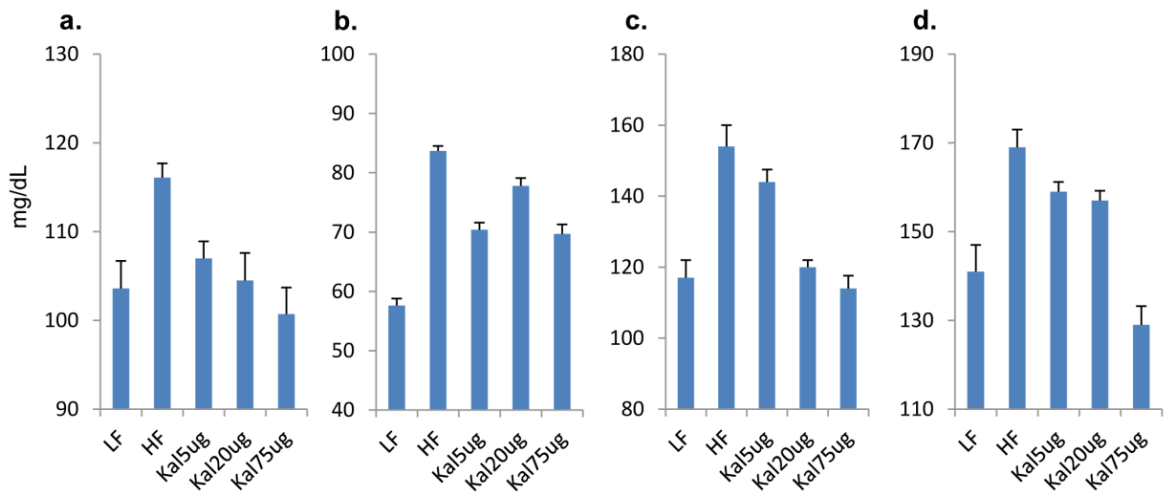

**Figure 2**

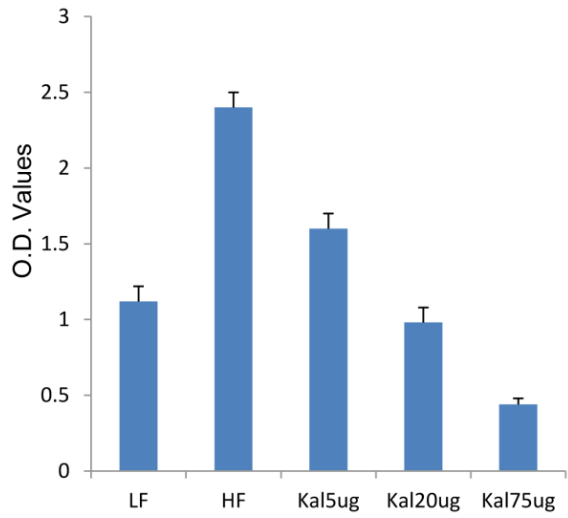
**Figure 3A**

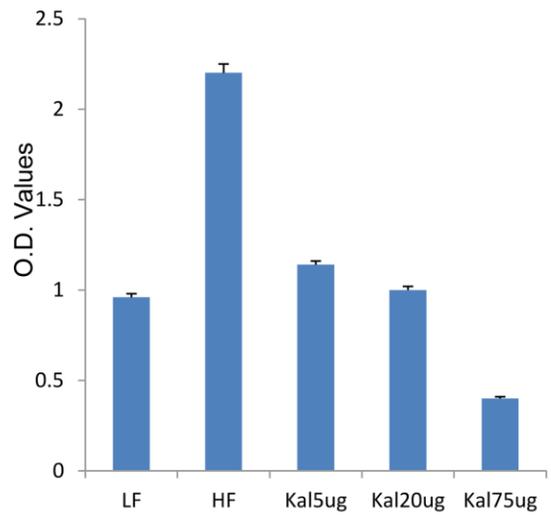
**Figure 3B**

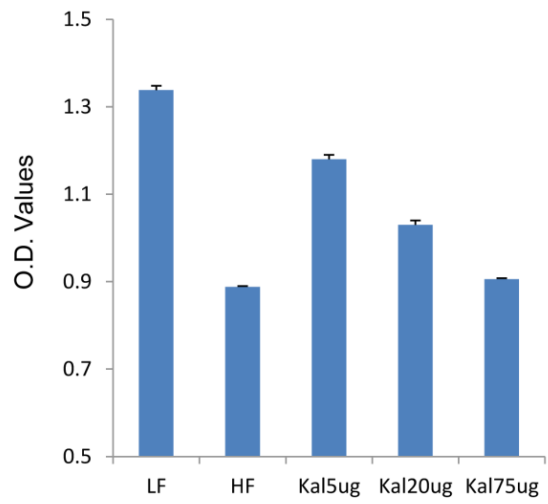
**Figure 3C**

**Figure 3**

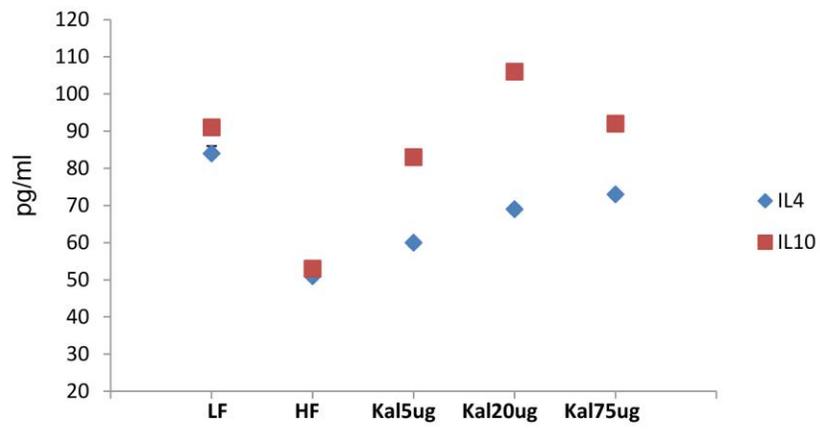

**Figure 4A**

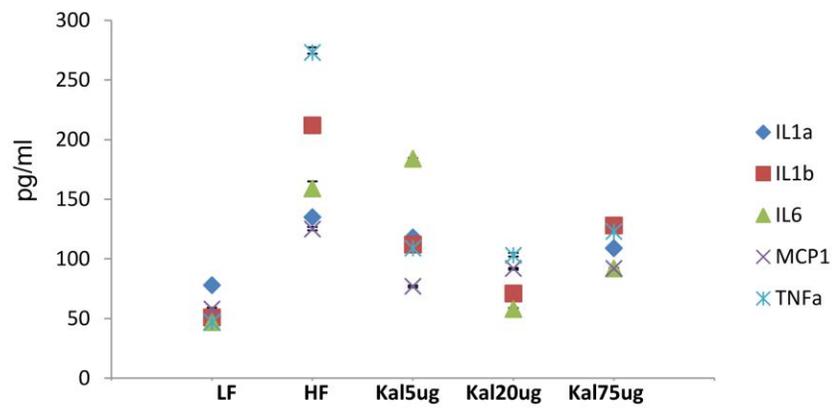

**Figure 4B**

**Figure 4**

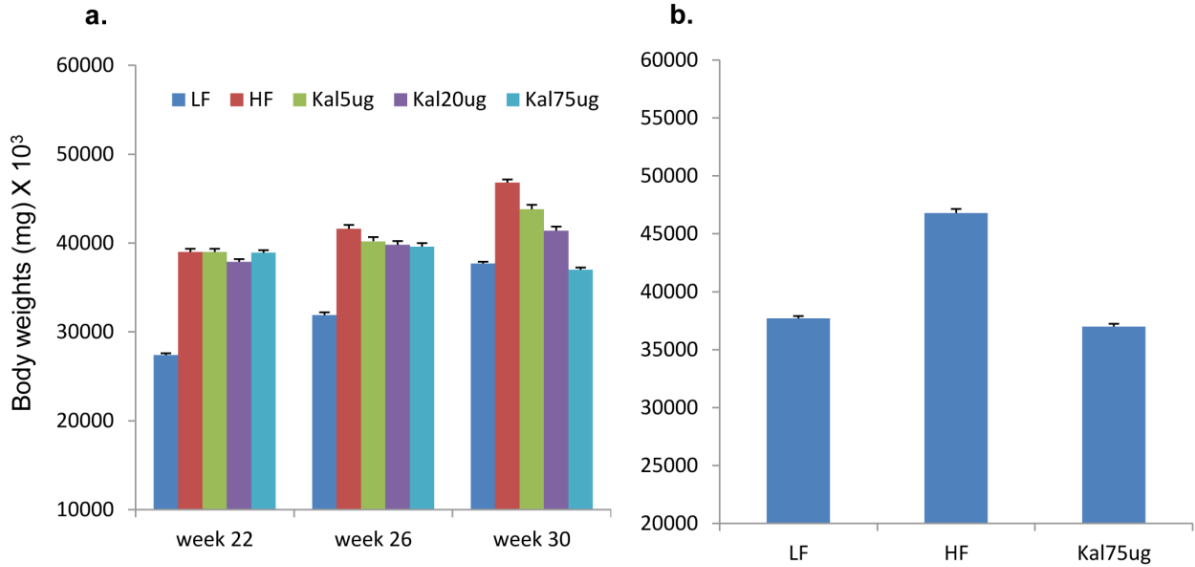

**Figure 5A**

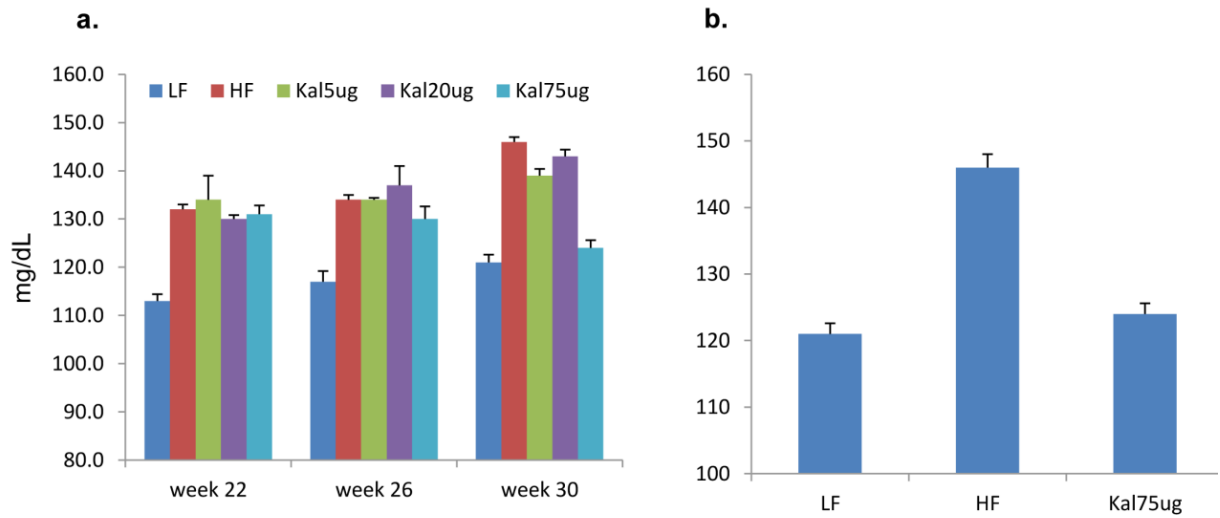

**Figure 5B**

**Figure 5**

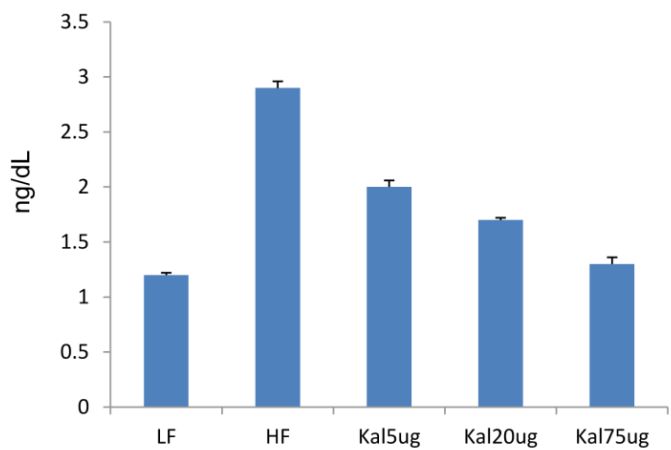

**Figure 5C**

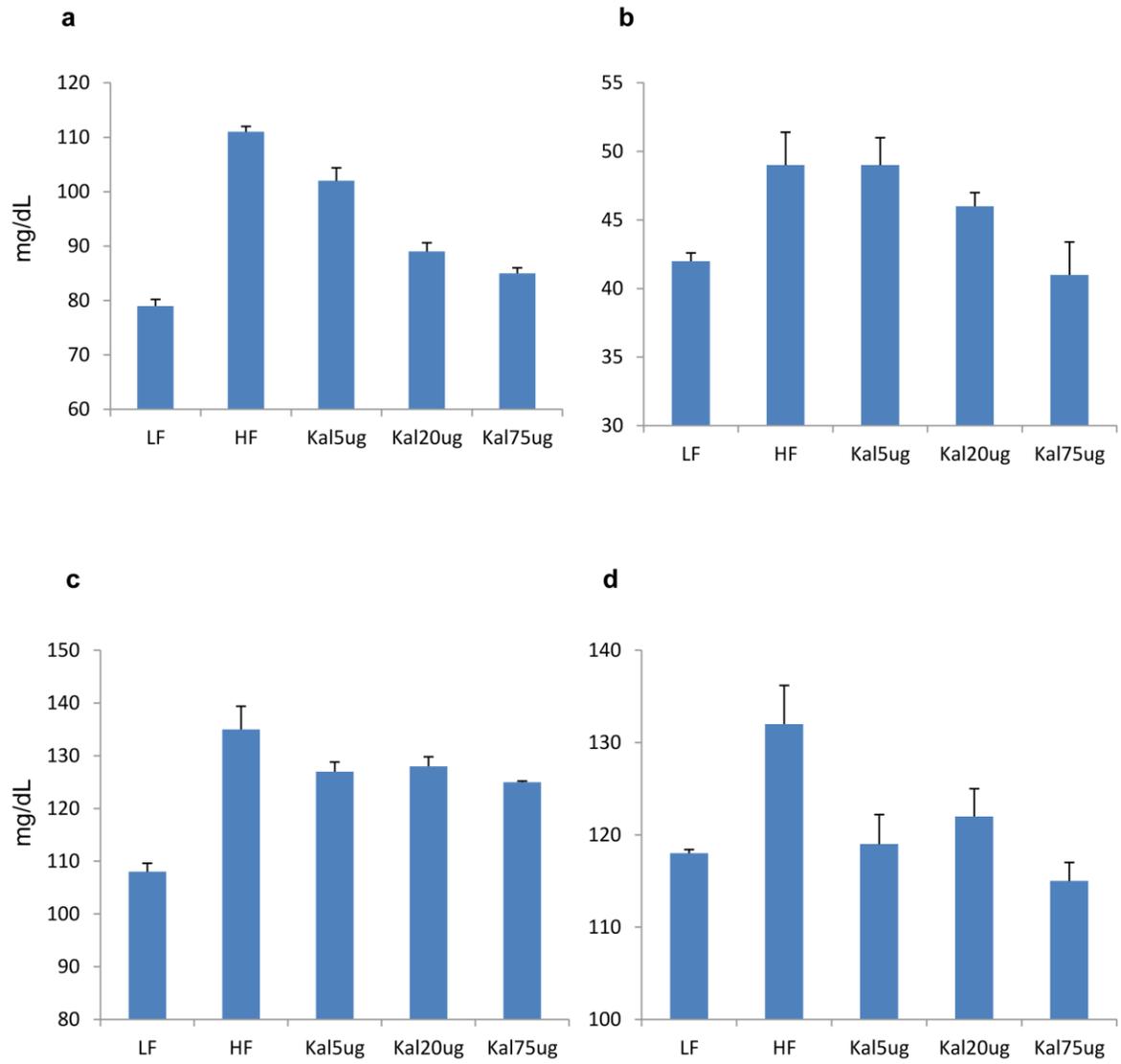

**Figure 6**

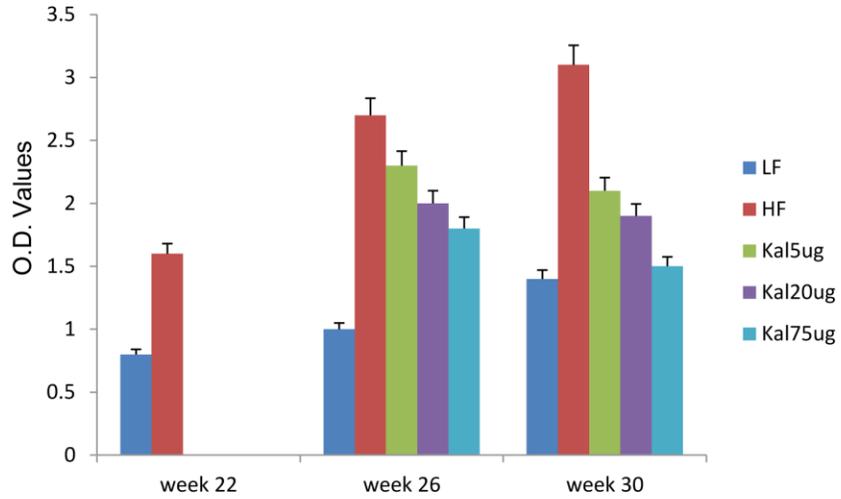

**Figure 7A**

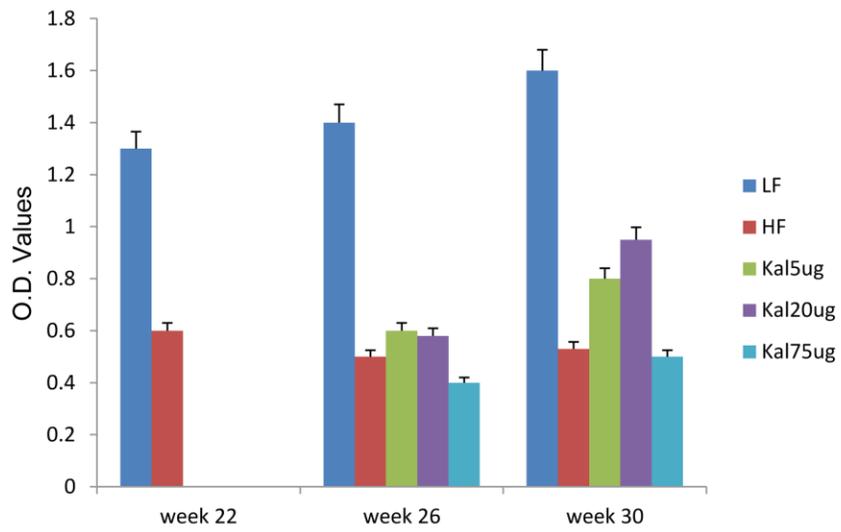

**Figure 7B**

**Figure 7**

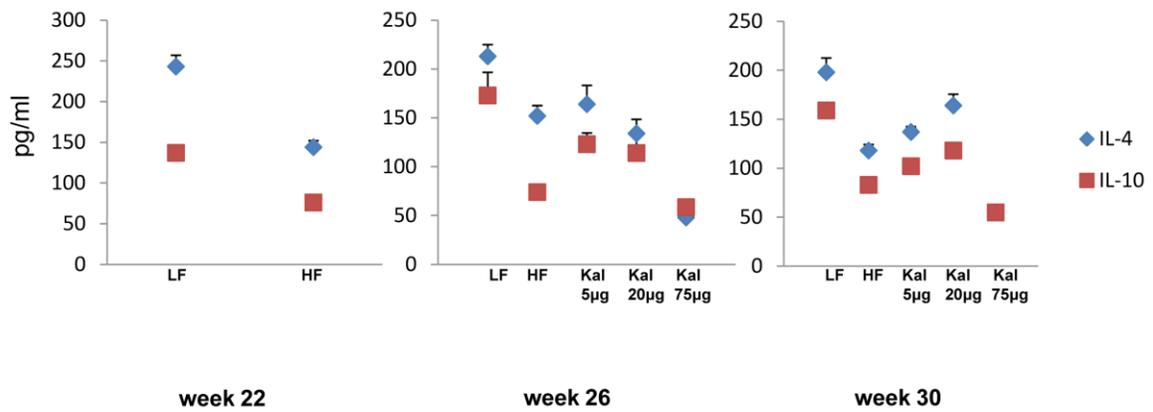

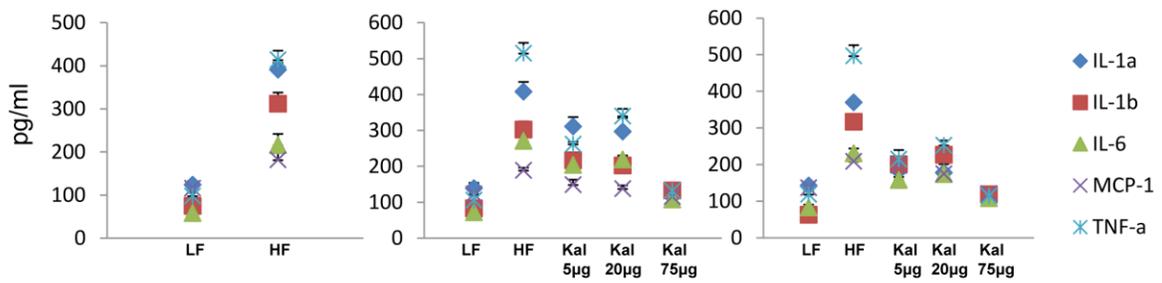

**Figure 8**

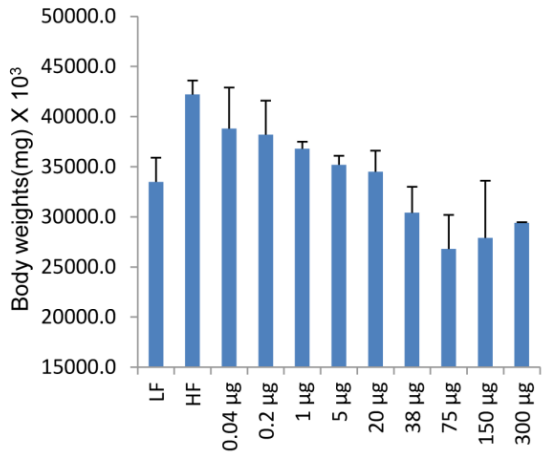

**Figure S1**

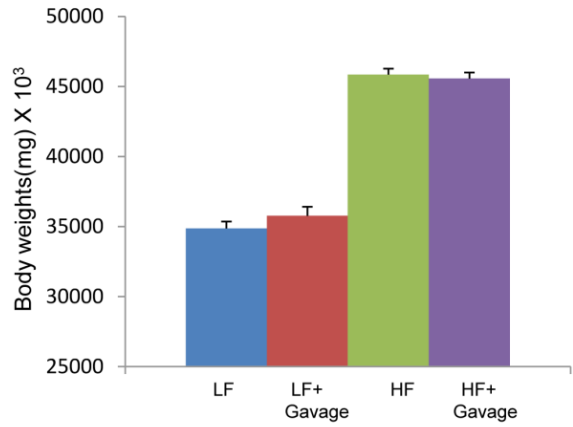

**Figure S2**

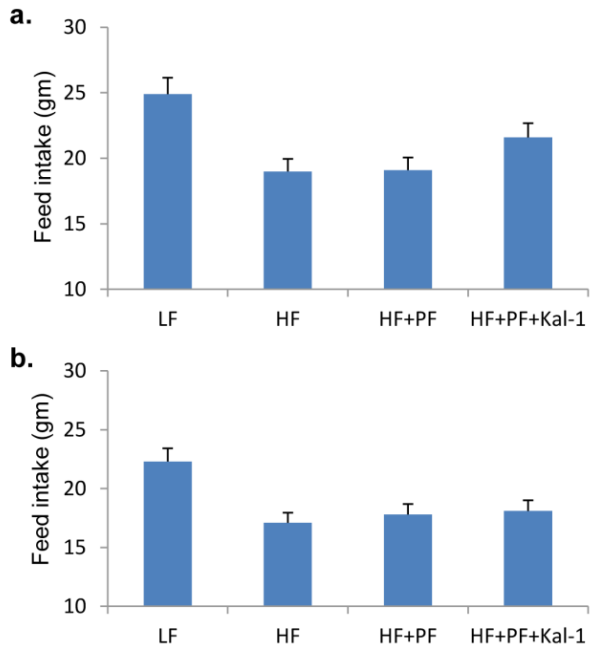

**Figure S3**

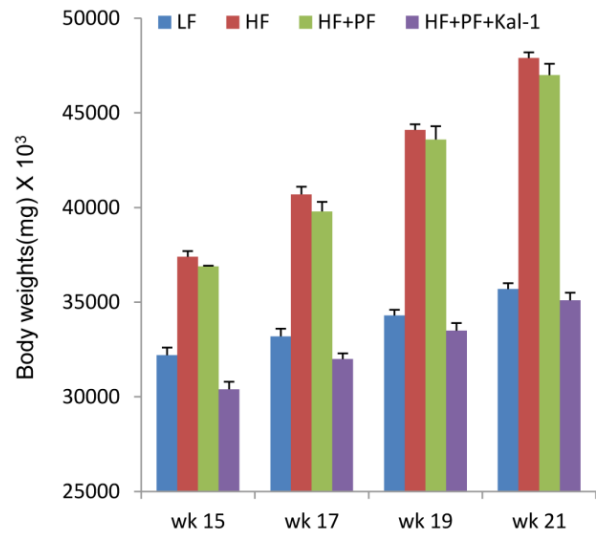

**Figure S4**

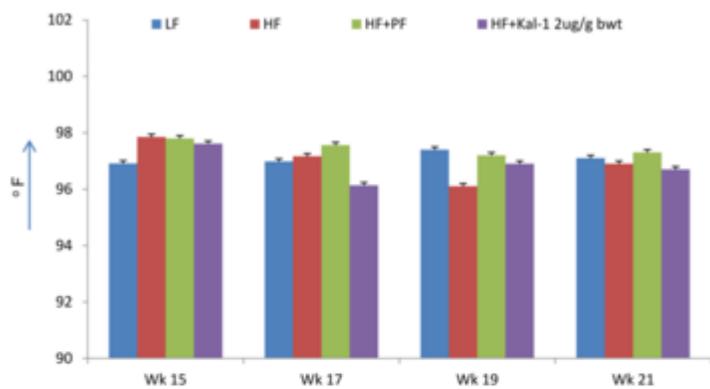

Figure S5

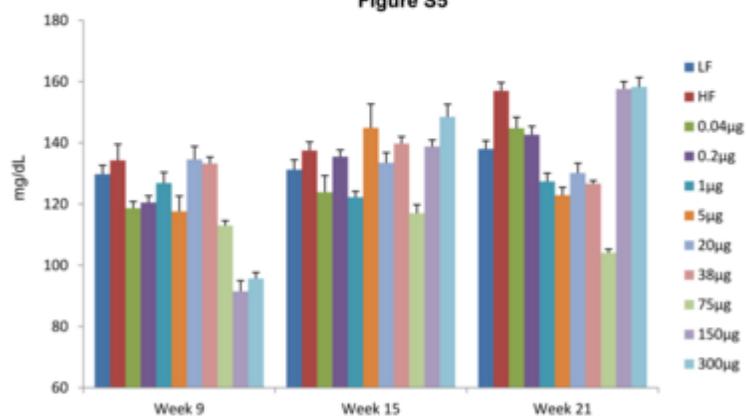

Figure S6

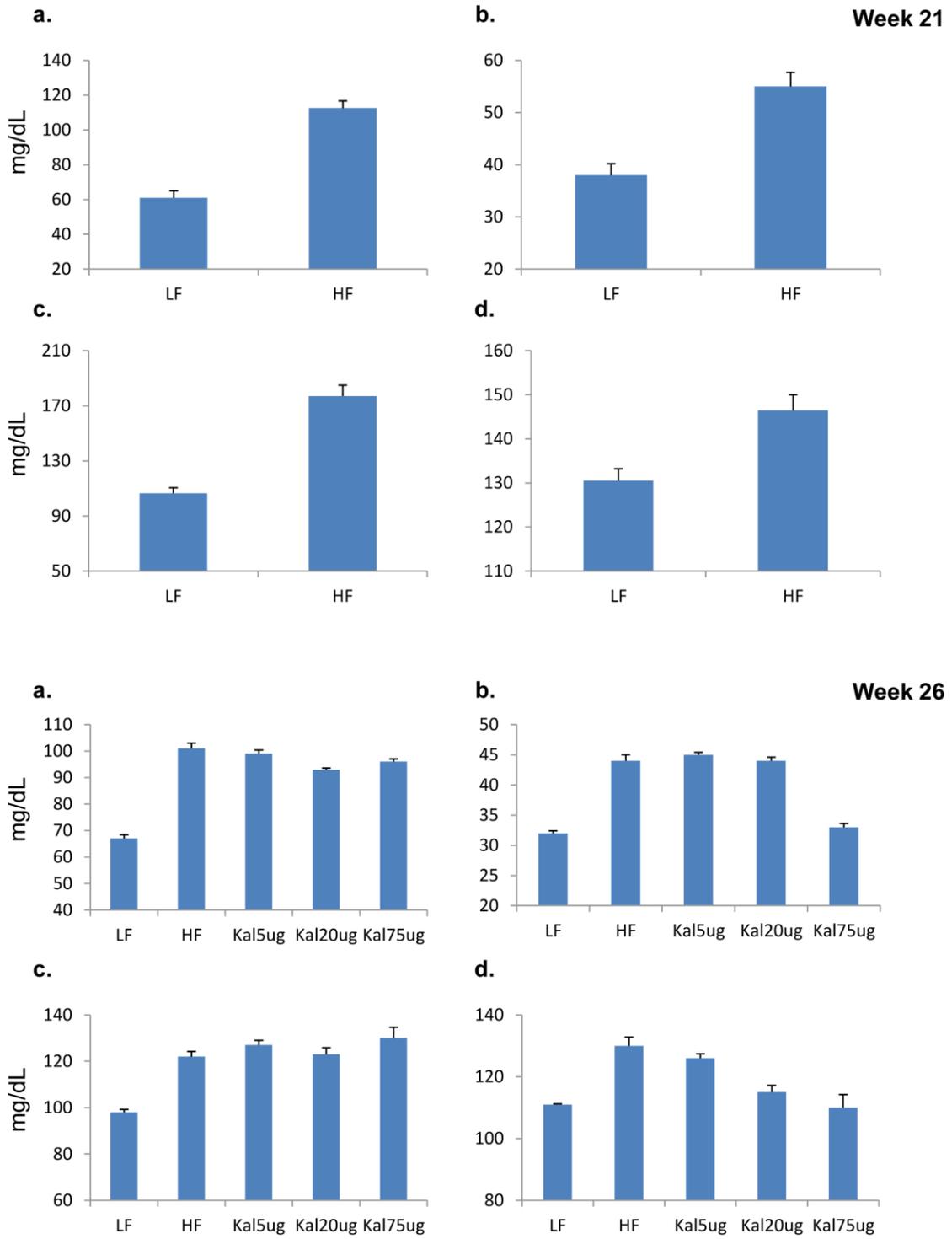

**Figure S7**

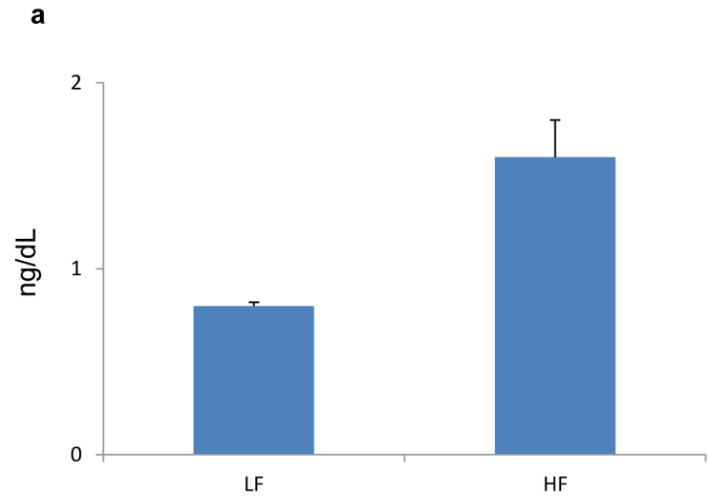

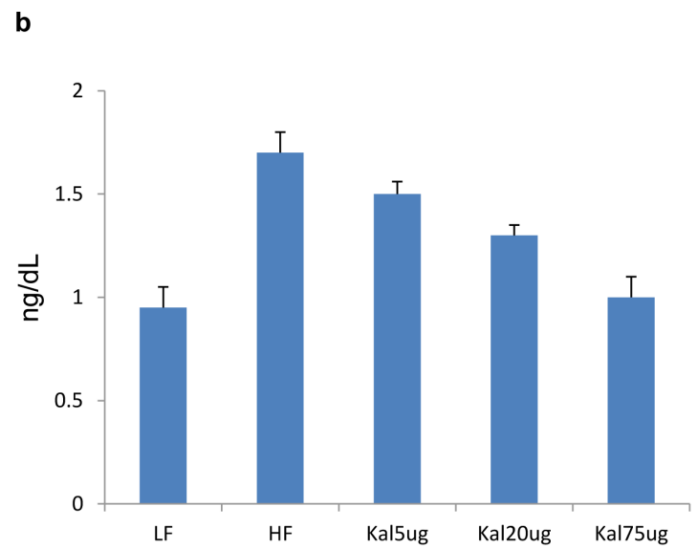

**Figure S8**

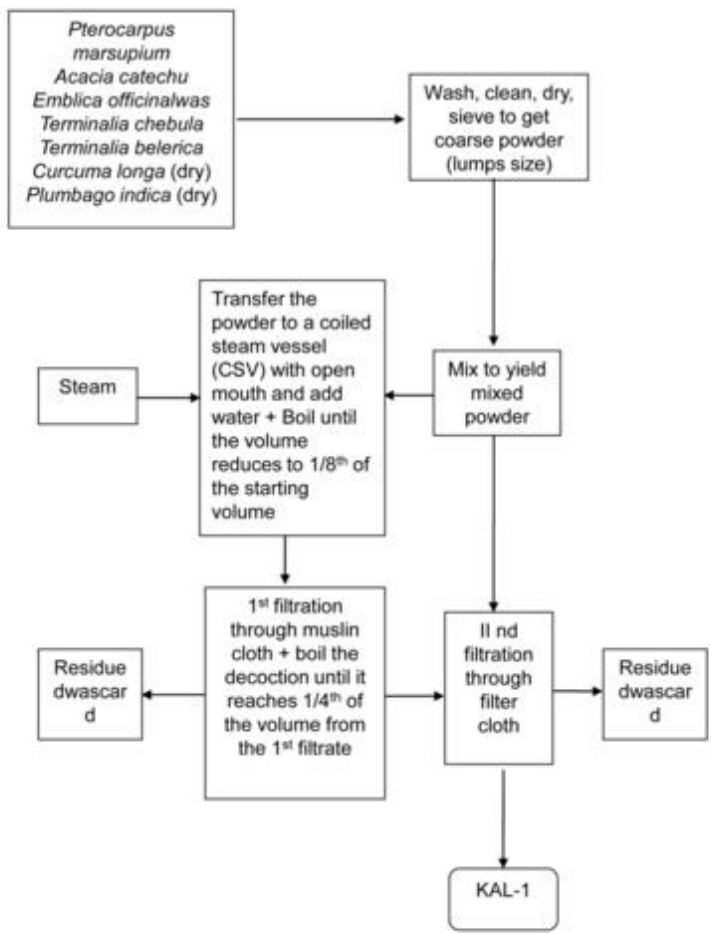

Figure S9